\documentstyle[12pt,epsf,psfig]{article}
\let\section=\subsection     \let\subsection=\subsubsection                
\newcommand{\be}{ \begin{eqnarray}}
\newcommand{\ee}{\end{eqnarray}}

\begin{document}
\hfill LBNL-45200\\
\ \\

\begin{center}
   {\large \bf DILEPTON PRODUCTION AND CHIRAL SYMMETRY}\\[2mm]
   V.~KOCH$^1$, M.~BLEICHER$^1$, A.K.~DUTT-MAZUMDER$^2$, 
    C.~GALE$^2$, C.M.~KO$^3$ \\[5mm]
   {\small \it (1) Lawrence Berkeley National Laboratory, Berkeley, CA 94720\\
               (2) McGill University, Montreal, Quebec H3A, Canada\\
               (3) Texas A\&M University, College Station, TX 77843, USA
                \\[8mm] }
\end{center}

\begin{abstract}\noindent
   We discuss how dilepton production is related to  chiral symmetry
   and its restoration. We then analyse presently available data
   by the CERES collaboration in this context. We find that the present data
   do not support any conclusions concerning the restoration of chiral
   symmetry. We finally provide a prediction for the dilepton spectrum for the
   just completed low energy (40~GeV) run at the SPS. 
\end{abstract}

\section{Introduction}
Electromagnetic probes provide a unique tool to investigate the early and
interior properties of the system created in a heavy ion collision. Dileptons
are of particular interest as their production cross section is related to the
correlation function of the iso-vector vector current, the conserved current of
the $\rm SU_R(2) \times SU_L(2)$ chiral symmetry. Thus, a careful study of the
dilepton spectrum may provide some evidence and signal for the existence of a
chirally restored phase created in a heavy ion collisions. Of  course,
unambiguous proof of chiral restoration requires the measurement of both
iso-vector vector and iso-vector axial-vector correlation functions. The
detection of the latter, however, is rather difficult, as it does not couple
to a penetrating probe but rather to pionic degrees of freedom, and thus is
strongly affected by final state interactions. However, at least at low
temperatures one can show \cite{ioffe}, that the onset of chiral restoration
goes via the mixing of the 
vector and axial correlators. Assuming that this mixing
is the dominant mechanism for chiral restoration (for a discussion of other
possibilities see e.g. \cite{kapusta}), one expects considerable changes in the
vector-correlator, which is measurable via dilepton production.

Let us, however, already note at this point that the medium can not only give
rise to mixing of vector and axial vector correlation functions in the
iso-vector channel, but also to mixing between iso-scalar and iso-vector 
(`$\rho$ and $\omega$') correlator. This mixing is {\em not} related to
chiral restoration, but certainly contributes to the dilepton spectrum, as
both iso-vector and iso-scalar currents couple to the electromagnetic field. 
For the purpose of extracting information about possible chiral restoration,
it has to be considered as `background'.       

\section{Chiral restoration and mixing}
As already mentioned in the introduction, to leading order in the temperature 
the iso-vector vector correlator receives an admixture from the axial
correlator \cite{ioffe}. Similar arguments, although somewhat model dependent,
can also be given at finite density, 
where one again finds this mixing of vector
and axial vector correlator \cite{chanfray}. In more physical terms, in both
cases a pion from either the heat bath (finite temperature) or from the pion
could of the nucleons (finite density) couples to the vector-current ($\rho$ 
meson) to form an axial-vector ($a_1$ or pion) intermediate state. This is
depicted schematically on the left hand side of  fig. 1. Conversely, the
pions from the heatbath and/or pion cloud also give rise to a
vector-admixture to the axial-vector correlator. One, therefore,  
can imagine that a
certain temperature/density vector and axial-vector correlators are fully
mixed and thus indistinguishable, which in turn means that chiral symmetry is
restored. 

The pions in the medium, however, also give rise to other mixing. For example
a pion from the heatbath may couple to a $\rho$-meson to give an $\omega$
intermediate state. This corresponds to the mixing between the iso-vector
vector and iso-scalar vector correlators. This process is depicted
schematically on the right hand side of fig. 1. This mixing is {\em not} 
related to chiral symmetry restoration, but still affects the dilepton
spectrum. 

The contribution of these mixings to the dilepton spectrum can be easily
understood in simple physical terms. The dilepton-production cross
section is proportional to the imaginary part of these correlation functions,
which corresponds to a simple cut of the diagrams depicted in fig.1. Therefore,
the matrix-elements responsible for vector axial-vector mixing are identical
to those for the Dalitz-decay of the
$a_1$-meson. A similar relation hold between the the mixing of iso-vector
and iso-scalar correlators to the Dalitz decay of the $\omega$ meson. 

In other words, the contribution of the $a_1$-Dalitz decay to the total
dilepton spectrum provides a measure for the importance of the vector
axial-vector mixing, i.e. for the sensitivity to chiral restoration effects.  
While it might be impossible to extract this channel from the data, in a given
model calculation it can be easily accessed. We will show below, that the
$a_1$-Dalitz contribution is so small that its absence or presence would make
an indistinguishable difference in the total spectrum. Consequently, the
sensitivity to chiral restoration appears to be very weak.

\begin{center}
\begin{minipage}{13cm}
\baselineskip=12pt
\epsfxsize=.8 \textwidth
\par
\centerline{\epsffile{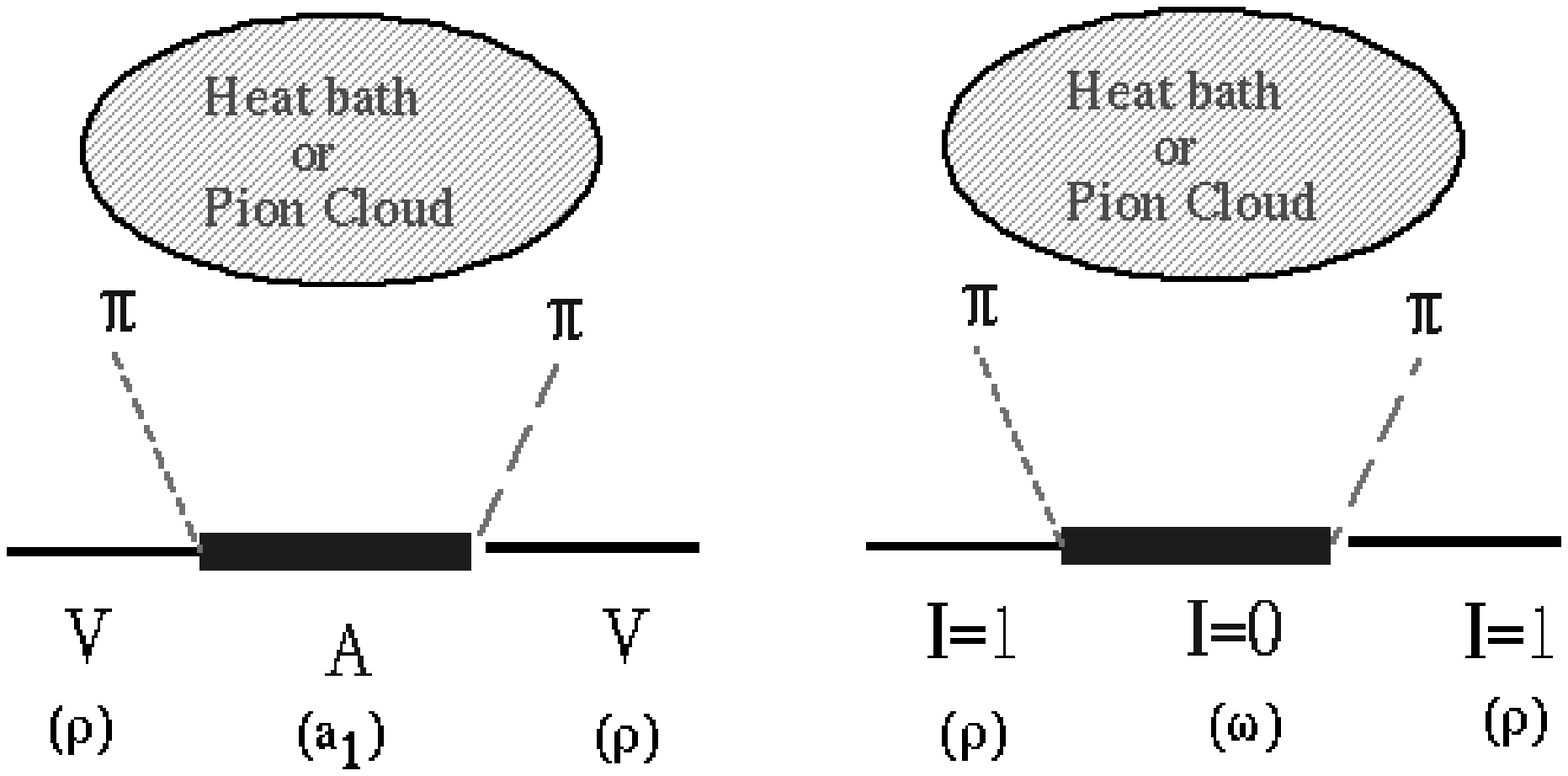}}

{\begin{small}
Fig.~1. Schematic illustration of mixing induced by the medium. Left: Vector
axial-vector mixing. Right: Iso-vector iso-scalar mixing.\end{small}}
\end{minipage}
\end{center}

Finally, there has been discussion about the contributions of baryons, notably
the $N(1520)$ resonance \cite{Mosel,rapp}. Again, this can be understood
either as a mixing of the $\rho$-meson with $N(1520)$-hole states or simply as
the Dalitz decay of the $N(1520)$. At present, it is unclear what the chiral
properties of a $N(1520)$-hole state are. Therefore, one cannot say to which
extent this mixing contributes to chiral restoration. 
However, as will be shown below, the contribution of the $N(1520)$-Dalitz 
again is so small, that it hardly affects the total Dilepton spectrum.

\section{The N(1520)}
There has been some discussion as to how important the contribution of baryons
to the dilepton spectrum is. Baryon-resonances contribute chiefly through 
their Dalitz decay $N^*\rightarrow e^+ e^- N$ to the dilepton spectrum. In 
\cite{song} this contribution has been estimated to be at most 50 \% of that of
the $\omega$-Dalitz decay in the mass range of $400-500 \,\rm MeV$. The
$N(1520)$ plays a special role, as it couples very strongly to the $\rho$ meson
and thus should also contribute most to the dilepton spectrum 
\cite{Mosel,rapp}. In order to explore this we have calculated the Dalitz
decay rate and width of the $N(1520)$ using the model and parameters of
\cite{feuster}, which is based on an analysis of pion photoproduction data. 
A detailed description of this calculation can be found in \cite{Abhee}. We
have also compared to the nonrelativistic models, which are usually employed
in the calculation of the $\rho$-meson spectral function \cite{Mosel,rapp}.  
In fig.~2 we show the resulting width for two relativistic parameterizations
and the nonrelativistic result. They agree reasonably well, and in the
transport calculations shown below we employ the nonrelativistic result. 
We should also mention that we found a rather weak dependence on the off-shell
parameter, which is always present in the relativistic description of a spin
3/2 object.

\begin{center}
\begin{minipage}{13cm}
\baselineskip=12pt
\epsfysize=7cm
\par
\centerline{\epsffile{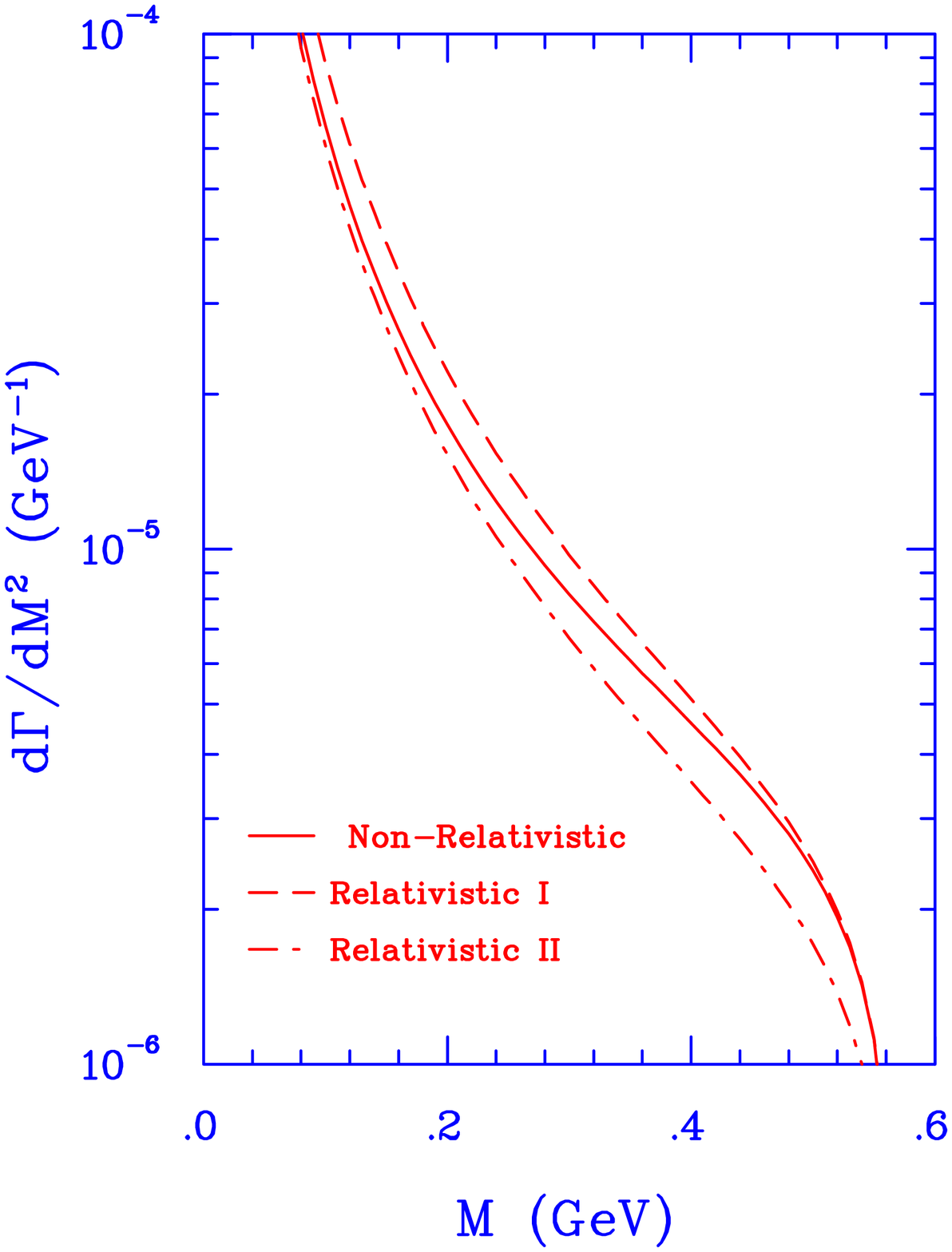}}

{\begin{small}
Fig.~2. Dalitz decay width of the N(1520) based on a relativistic and
nonrelativistic descriptions. The difference between the calculations labeled
`Relativistic I' and `Relativistic II' are discussed in detail in
\protect\cite{Abhee}. \end{small}}
\end{minipage}
\end{center}

\section{Transport calculation}
The results which we will present in the subsequent section are based on a
transport calculation. Therefore, it is appropriate to discuss how the
dilepton spectrum calculated in transport is related to that obtained from in
medium correlation functions (see e.g. \cite{rapp}). In general, the dilepton
production rate is related to the imaginary part of the current-current
correlation function $C(q_0,\vec{q})$ 
\be
\frac{d R}{d^4 q} = - \frac{\alpha^2}{\pi^3 q^2} {\rm Im} \, 
C(q_0,\vec{q}). 
\ee
Assuming vector dominance, and concentrating on the iso-vector vector ($\rho$)
channel the current-current correlator is, up to constant factors, given by
the in medium rho propagator $D_\rho(q_0,\vec{q})$
\be
{\rm Im} \, C(q_0,\vec{q}) \sim {\rm Im} \, D\rho(q_0,\vec{q}) 
\sim \frac{{\rm Im} (\Sigma)}{(m^2 - m_\rho^2 + {\rm Re} \,( \Sigma))^2 +  
{\rm Im} \,( \Sigma)^2}
\label{corr_full}
\ee
with  
\be
\Sigma = \Sigma_{\rho\rightarrow \pi \pi} + \Sigma_{\pi+\rho \rightarrow a_1} 
+ \ldots
\ee
being the selfenergy of the $\rho$ meson. In free space $\Sigma =
\Sigma_{\rho\rightarrow \pi \pi}$. In the medium, one gets additional
contributions from all the other channels coupling to the $\rho$ such as the 
$a_1$, the $\omega$, $N(1520)$ etc. 

In transport, on the other hand, one folds the collisions of the respective
particles with the branching probability into the dilepton channel in order to
generate the dilepton spectrum. 
At first sight, this appears to be a different approach then the one based on
the imaginary part of the correlator. However, 
the relevant cross sections which control the
collision probabilities are directly related to the imaginary part of the
selfenergies. Thus the transport calculation can also be formulated in terms
of an imaginary part of a correlation function which, for the results shown
below has the following form
\be
{\rm Im} \, C(q_0,\vec{q})_{transport} \sim {\rm Im} \, D\rho(q_0,\vec{q})_{transport} 
\sim \frac{{\rm Im} (\Sigma)}{(m^2 - m_\rho^2 )^2 +  
{\rm Im} \,( \Sigma_{\rho \rightarrow \pi \pi})^2}.
\label{corr_trans}
\ee
Note that there are two differences between the correlator (\ref{corr_full}) 
and 
that resulting from transport (\ref{corr_trans}). First, in the denominator the
real part of the selfenergy in neglected. This is a reasonable approximation
as the real part appears to be small \cite{rapp}. The other approximation is
that in the denominator only  the free selfenergy enters in the imaginary
part. Additional broadening due to other processes is not taken into
account. This approximation becomes bad close to the resonance mass, where the
imaginary part of the selfenergy dominates the denominator. There, the
transport over predicts the actual dilepton yield. However, in the results
below we will see that in this region the total dilepton spectrum is dominated
by the decay of the $\omega$ meson, which happens mostly in the final
state. Thus, a reduction of the $\rho$ peak will reduce the overall strength
only by a small amount. This effect will become more visible, once the
experimental mass resolution becomes sufficiently good to resolve the omega
peak. Then the collisional broadening of the $\rho$ should be directly visible
in the neighborhood of the omega.
One should also note that this second approximation can in principle be 
overcome within a transport calculation. The transport gives the full
information about the collisional width ($Im(\Sigma)$)
for each spacetime point. It, however, requires a numerical tour de force to
do a statistically reliable calculations. While this has not yet been attempted
in the framework of nucleus-nucleus collisions, such a calculation has been 
carried out for dilepton production in photon-nucleus reactions \cite{effe}.

\section{Results for Pb+Au at 160 GeV/A and predictions for 40 GeV/A}
Let us now turn to the actual results. Most of the details of the calculations
can be found in \cite{song}. The new element here is the inclusion of the
N(1520). In fig.~3 and fig.~4 we show the resulting dilepton spectra in 
comparison with the preliminary CERES data from 1996 \cite{CERES}. We find an
overall agreement with the data, also for the low transverse momentum
ones. Notice, these results have been obtained without any in-medium effects. 
As explained in \cite{song} the initial conditions have been chosen such that
the final hadronic spectra agree with experiment. 

\begin{center}
\begin{minipage}{13cm}
\baselineskip=12pt
\epsfysize=8cm
\par
\centerline{\epsffile{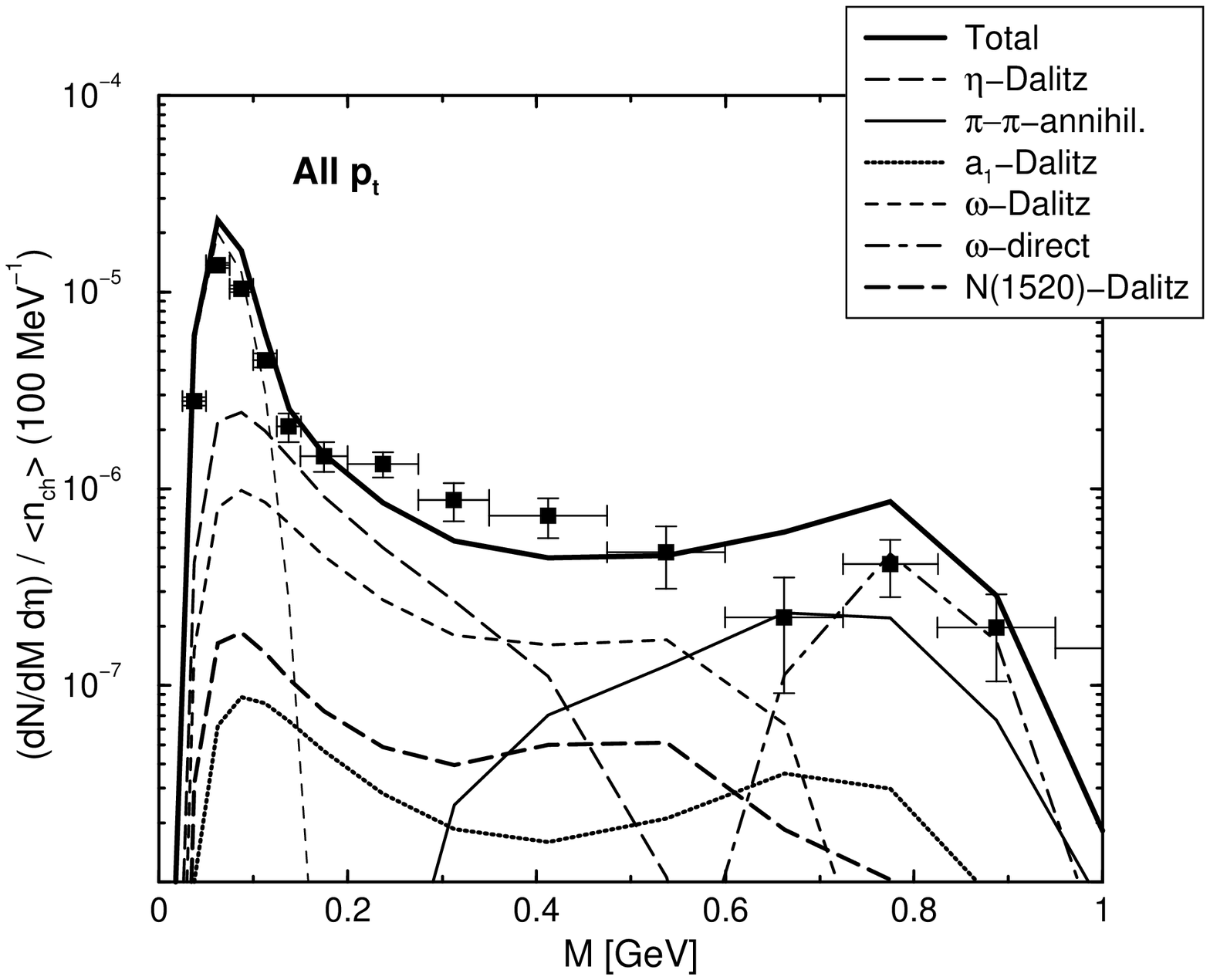}}

{\begin{small}
Fig.~3. Results for the dilepton invariant mass spectrum at 
$160 \,\rm GeV/A$. All transverse momenta. Calculation based on model of
\protect\cite{song}. \end{small}}
\end{minipage}
\end{center}

As discussed above, an indicator for the chiral mixing of axial-vector and
vector correlator is the $a_1$ Dalitz contribution (thick full line at the
bottom of the graph). In our calculation the $a_1$-Dalitz is at best a 2 \%
contribution. Considering the size of the experimental error bars, it is clear
that the present data do not allow for any conclusions concerning chiral
restoration. Note that our findings here are also in agreement with refs. 
\cite{haglin} and \cite{gale}. Both references 
find that the $a_1$-Dalitz rate is
considerably smaller than the pion annihilation rate.      

We also find a rather small contribution from the $N(1520)$ decay (thick
dashed line at the bottom of the graph). It is about a factor of 4 below the
contribution of the $\omega$-Dalitz decay, supporting the early estimate of
\cite{song}. This also implies that at least for the $160 \, \rm GeV$ data,
the contribution of the baryons is rather small.

\begin{center}
\begin{minipage}{13cm}
\baselineskip=12pt
\par
\centerline{\psfig{figure=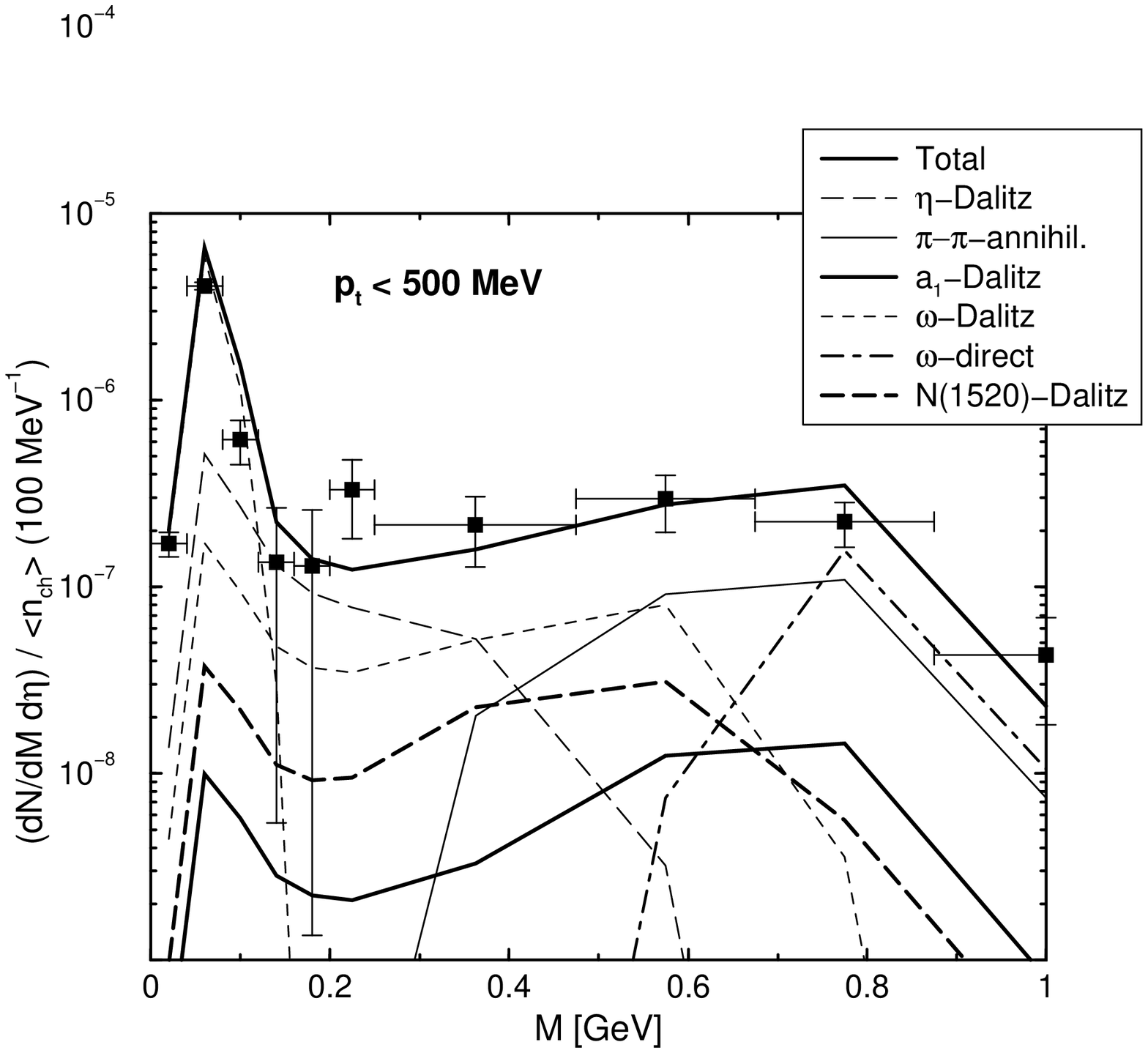,height=8cm,clip=}}

{\begin{small}
Fig.~4. Results for the dilepton invariant mass spectrum at 
$160 \,\rm GeV/A$. Transverse momenta below 500~MeV/c. 
Calculation based on model of
\protect\cite{song}.
\end{small}}
\end{minipage}
\end{center}

Let us next turn to our prediction for the `low energy' ($40 \,\rm GeV$) run. 
In figs.~5 and 6 we show our prediction together with the 160~GeV data, which
are intended as a reference. 
The prediction is based on central ($N_{charge} = 220$) 
events generated with URQMD \cite{urqmd}
(for details of the calculation see \cite{Abhee}). 
In our calculation, we have assumed a mass resolution of 1 \%, in order to
account for the substantially improved mass resolution of the CERES
spectrometer.  Given this mass resolution, the omega should be clearly
visible and thus should put a strong constraint on the model
calculations. Aside from the improved resolution, however, we do not predict a
significant difference between the high energy and the low energy spectrum, if
plotted in the CERES normalization i.e. if divided by the number of charged
particles. The only small but visible difference is around 400~MeV, where 
the yield is somewhat smaller than that of the high energy run.
Finally, we find that the contribution of the 
baryons is still comparatively small,
although somewhat stronger than for  the high energy collisions. 

We have also carried out a calculation in the spirit of
\cite{song}, where we have adjusted the initial hadronic state in order to
reproduce the final spectra of an RQMD calculation at 40~GeV/A. The results
are virtually identical to the ones using the full URQMD events.

\begin{center}
\begin{minipage}{13cm}
\baselineskip=12pt
\epsfysize=8cm
\par
\centerline{\epsffile{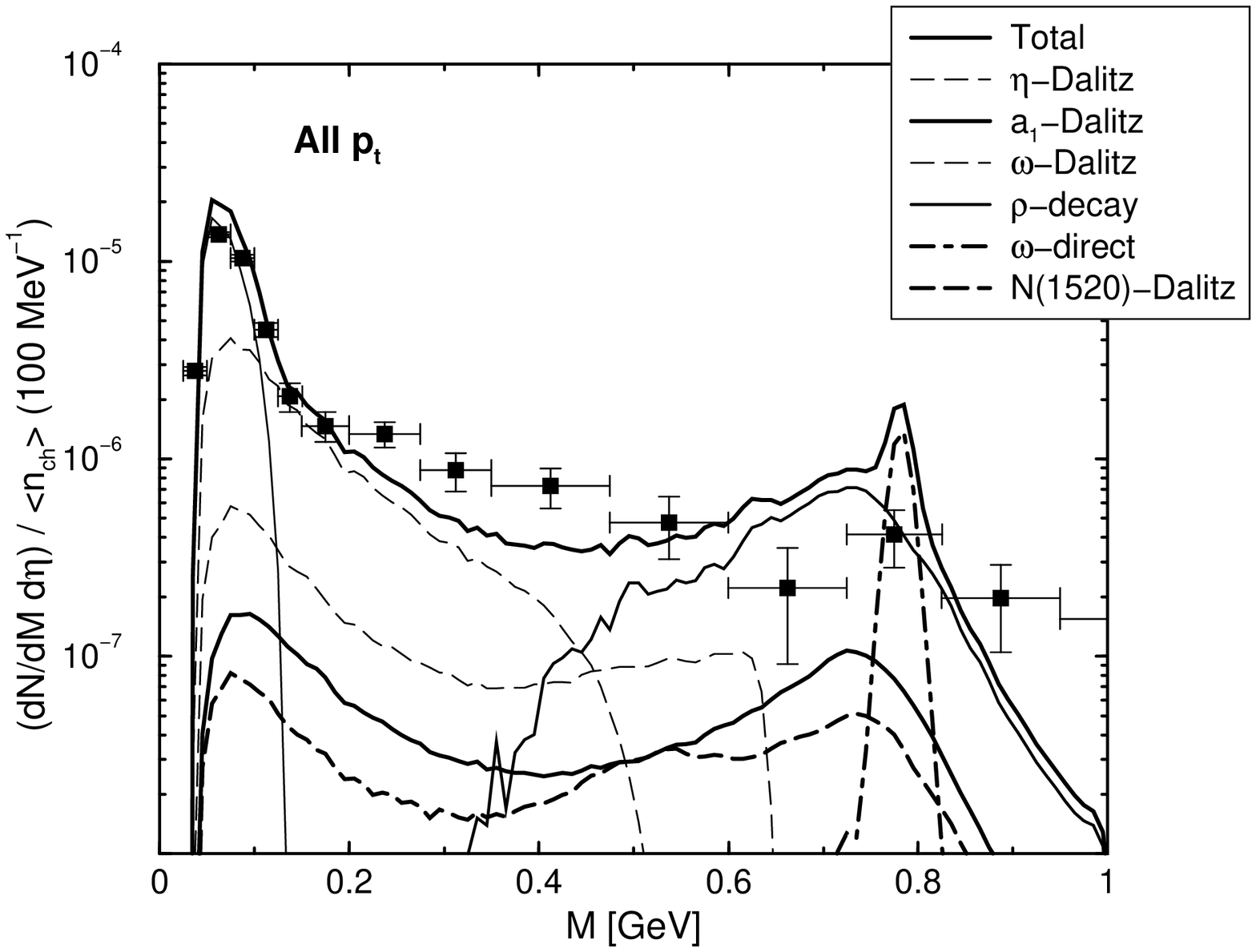}}
{\begin{small}
Fig.~5. Prediction for the dilepton invariant mass spectrum for Pb+Pb at 
$40 \,\rm GeV/A$. All transverse momenta. Calculation based on URQMD
\protect\cite{urqmd}. The data are those for 160~GeV/A \protect\cite{CERES}
and shown for comparison only. 
\end{small}}
\end{minipage}
\end{center}

\section{Conclusions}
The imaginary part of the iso-vector vector correlator contributes
significantly  to the dilepton spectrum. Therefore, a dilepton measurement is
in principle sensitive to in medium changes of this correlation function, and
thus may be utilized to investigate effects of chiral symmetry restoration,
at least indirectly. We have argued that the strength of the $a_1$-Dalitz decay
provides a good measure for the mixing of the vector and axial-vector 
correlators, which is one way of restoring chiral symmetry. Our 
analysis of the presently available dilepton data by the CERES collaboration
shows a negligible contribution from the $a_1$-Dalitz decay, 
in agreement with other calculations. Therefore, the present dilepton data 
do not allow for any
conclusions concerning the mixing of the axial-vector and vector correlation
functions, and thus about the possible restoration of chiral symmetry achieved
in SPS-energy heavy ion collisions. 

\begin{center}
\begin{minipage}{13cm}
\baselineskip=12pt
\epsfysize=8cm
\par
\centerline{\epsffile{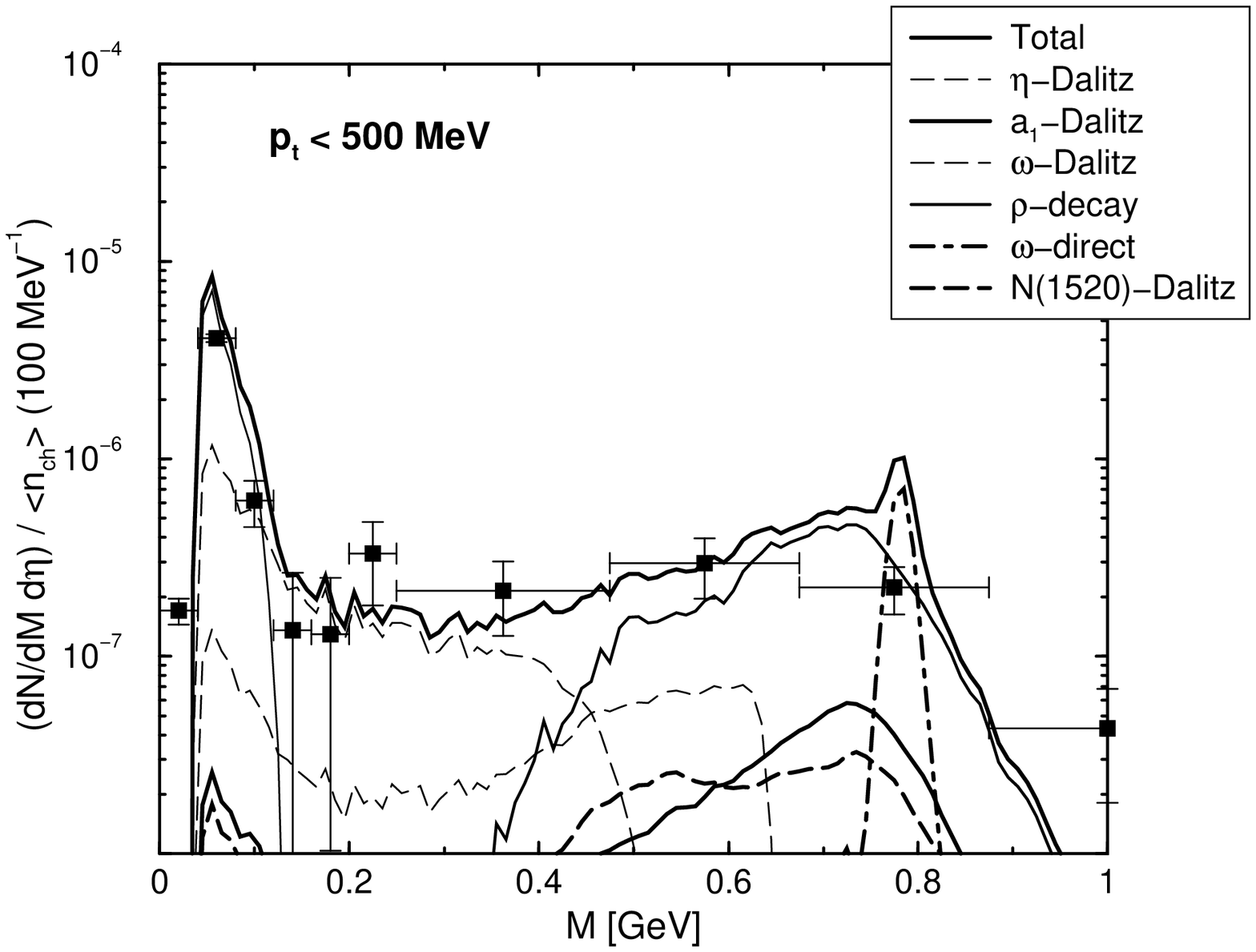}}
{\begin{small}
Fig.~6. Prediction for the dilepton invariant mass spectrum for Pb+Pb at 
$40 \,\rm GeV/A$. Transverse momenta below 500~MeV/c. 
Calculation based on URQMD \protect\cite{urqmd}.
The data are those for 160~GeV/A \protect\cite{CERES}
and shown for comparison only.
 \end{small}}
\end{minipage}
\end{center}

We have also provided a prediction for the dilepton spectrum in central Pb+Pb
collisions at 40~GeV/A. If normalized by the number of charged particles we
find only small differences between the low energy and the high energy
spectra. Furthermore, even at the low bombarding energy, we find that the
contribution from baryons is rather small.      

\ \\
\noindent
Acknowledgments: The work of CG and AKDM was supported
by the Natural Sciences and Engineering Research Council of Canada
and the Fonds FCAR of the Qu\'ebec Government. 
The work of CMK was
supported in part by the National Science Foundation under Grant
No. PHY-9509266 and PHY-9870038, the Welch Foundation under Grant
No. A-1358, the Texas Advanced Research Program under Grant FY97-010366-068, 
and the Alexander v.
Humboldt Foundation. MB and VK were supported by the Director, Office of Science, Office of High Energy and Nuclear Physics, 
Division of Nuclear Physics, and by the Office of Basic Energy
Sciences, Division of Nuclear Sciences, of the U.S. Department of Energy 
under Contract No. DE-AC03-76SF00098. MB was also supported by a Feodor Lynen
Fellowship of the Alexander v. Humboldt Foundation, Germany.

\end{document}